\documentclass[preprint]{revtex4}
\usepackage{graphicx}
\usepackage{amssymb,amsfonts,amsmath}
\usepackage{bm}
\newcommand{\comment}[1]{}

\begin{document}


\title{Stochastic analysis of a miRNA-protein toggle switch}
\author{E. Giampieri, D. Remondini, L. de Oliveira,G. Castellani}
\affiliation{Dip. di Fisica, Universit\'a di Bologna, Viale B. Pichat 6/2, 40126 Bologna IT}  
\author{P. Li\'o}
\affiliation{ Computer Science dept, University of Cambridge, JJ Thomson rd, Cambridge, UK} 
\maketitle

\section*{Abstract}
Within systems biology there is an increasing interest in the stochastic behavior of genetic and biochemical reaction networks. An appropriate stochastic description is provided by the chemical master equation, which represents a continuous time Markov chain (CTMC). In this paper we consider the stochastic properties of a biochemical circuit, known to control eukaryotic cell cycle and possibly involved in oncogenesis, recently proposed in the literature within a deterministic framework.
Due to the inherent stochasticity of biochemical processes and the small number of molecules involved, the stochastic approach should be more correct in describing the real system: we study the agreement between the two approaches by exploring the system parameter space.  We address the problem by proposing a simplified version of the model that allows analytical treatment, and by performing numerical simulations for the full model. We observed optimal agreement between the stochastic and the deterministic description of the circuit in a large range of parameters, but some substantial differences arise in at least two cases: 1) when the deterministic system is in the proximity of a transition from a monostable to a bistable configuration, and 2) when bistability (in the deterministic system) is ''masked'' in the stochastic system by the distribution tails. The approach provides interesting estimates of the optimal number of molecules involved in the toggle. Our discussion of the points of strengths, potentiality and weakness of the chemical master equation in systems biology and the differences with respect to deterministic modeling are leveraged in order to provide useful advice for both the bioinformatician practitioner and the theoretical scientist.

Keywords: Toggle switch, chemical master equation, stochastic versus deterministic, biochemical reactions

\section*{Introduction}

Complex cellular responses are often modeled as switching between phenotype states, and despite the large body of deterministic studies and the increasing work aimed to elucidate the effect of intrinsic and extrinsic noise in such systems, some aspects still remain unclear. Molecular noise, which arises from the randomness of the discrete events in the cell (for example DNA mutations and repair) and experimental studies have reported the presence of stochastic mechanisms in cellular processes such as  gene expression \cite{McAdams},\cite{Elowitz}, \cite{Kepler}, decisions of the cell fate \cite{Arkin2}, and circadian oscillations \cite{Barkai}. Particularly, low copy numbers of important cellular components and molecules give rise to stochasticity in gene expression and protein synthesis, and it is a fundamental aspect to be taken into account for studying such biochemical models \cite{,Arkin,Cast09}. In this paper, we consider a simplified circuit that is known to govern a fundamental step during the eukaryotic cell cycle that defines cell fate, previously studied by means of a deterministic modeling approach \cite{Aguda}. 
Let set the scene by reminding that ''all models are wrong, but some are useful'' (said by George Edward Pelham Box, who was the son-in-law of Ronald Fisher). Biologists make use of qualitative models through graphs; quantitative modeling in biochemistry has been mainly based on the Law of Mass Action which has been used to frame the entire kinetic modeling of biochemical reactions for individual enzymes and for enzymatic reaction network systems \cite{Lund}. The state of the system at any particular instant is therefore regarded as a vector (or list) of amounts or concentrations and the changes in amount or concentration are assumed to occur by a continuous and deterministic process that is computed using the ordinary differential equation (ODE) approach. However, the theory based  on the Law of Mass Action does not consider the effect of fluctuations.
If the concentrations of the molecules is not great(on the order of Avogadro's number) we cannot ignore fluctuations. Moreover,  biological systems also show heterogeneity which occurs as a phenotypic consequence for a cell population given stochastic single-cell dynamics (when the population is not isogenic and in the same conditions). From a practical point of view, for concentrations greater than about 10 nM, we are safe using ODEs; considering a cell with a volume of $10^{-13}$ liters this corresponds to thousands of molecules that, under poissonian hypothesis, has an uncertainty in the order of $1\%$. If the total number of molecules of any particular substance, say, a transcription factor, is less than 1,000, then a stochastic differential equation or a Monte Carlo model would be more appropriate.  Similarly to the deterministic case, only simple systems are analytically tractable in the stochastic approach, i.e. the full probability distribution for the state of the biological system over time can be calculated explicitly, becoming computationally infeasible for systems with distinct processes operating on different timescales.
An active area of research is represented by development of approximate stochastic simulation algorithms. As commented recently by Wilkinson the difference between an ‘approximate’ and ‘exact’ model is usually remarkably less than the difference between the ‘exact’ model and the real biological process \cite{Wilkinson}.
Given we can see this either as an unsatisfactorily state of art or as a promising advancement, we can summarise the methodological approaches as following. Biochemical networks 
have been modeled using differential equations when considering continuous variables changing with time, or stochastic differential equations (SDE) for the trajectories of continuous random variables changing with time, and using the Gillespie algorithm that could be thought as an algorithm for the trajectory of discrete random variables changing with time. For a random variable changing with time one can either characterize it by its stochastic trajectories, as by the SDE and the Gillespie algorithm, or one can characterize its probability distribution as a function of time. The corresponding equation for the SDE is the Fokker-Planck equation, and the corresponding equation for the Gillespie algorithm is called the chemical master equation (CME) \cite{Liang}. 

Therefore, the chemical master equation is a Markov chain in the limit where time is continuous and it could be thought as the mesoscopic version of the Law of Mass Action, i.e. it extends the Law of Mass Action to the mesoscopic chemistry and biochemistry, see for example \cite{Qian,Wolf}.

Here we compare the results of a stochastic versus deterministic analysis of a microRNA-protein toggle switch involved in tumorigenesis with the aim of identifying the most meaningful amount of information to discriminate cancer and healthy states.
We show that the stochastic counterpart of such deterministic model has many commonalities with the deterministic one, but some differences arise, in particular regarding the number of stable states that can be explored by the system. The disagreement between the stochastic and deterministic description is observed in a ``ghost`` effect caused by the proximity to a deterministic bifurcation \cite{Strogatz}, and in a somehow opposite situation, in which the variance of the stable point can mask the detection of the second peak in the stationary distribution.   
In this paper we perform a numerical study of the complete two-dimensional model, but we consider also a simplified, biologically meaningful, version of the model that allows to calculate an exact solution, with a numerical characterization of the parameter ranges in which the two systems produce qualitatively similar results.

\section{Properties of a microRNA toggle switch}

\begin{figure}[ht]
\label{fig:comptog} \centering
\includegraphics[scale=0.8]{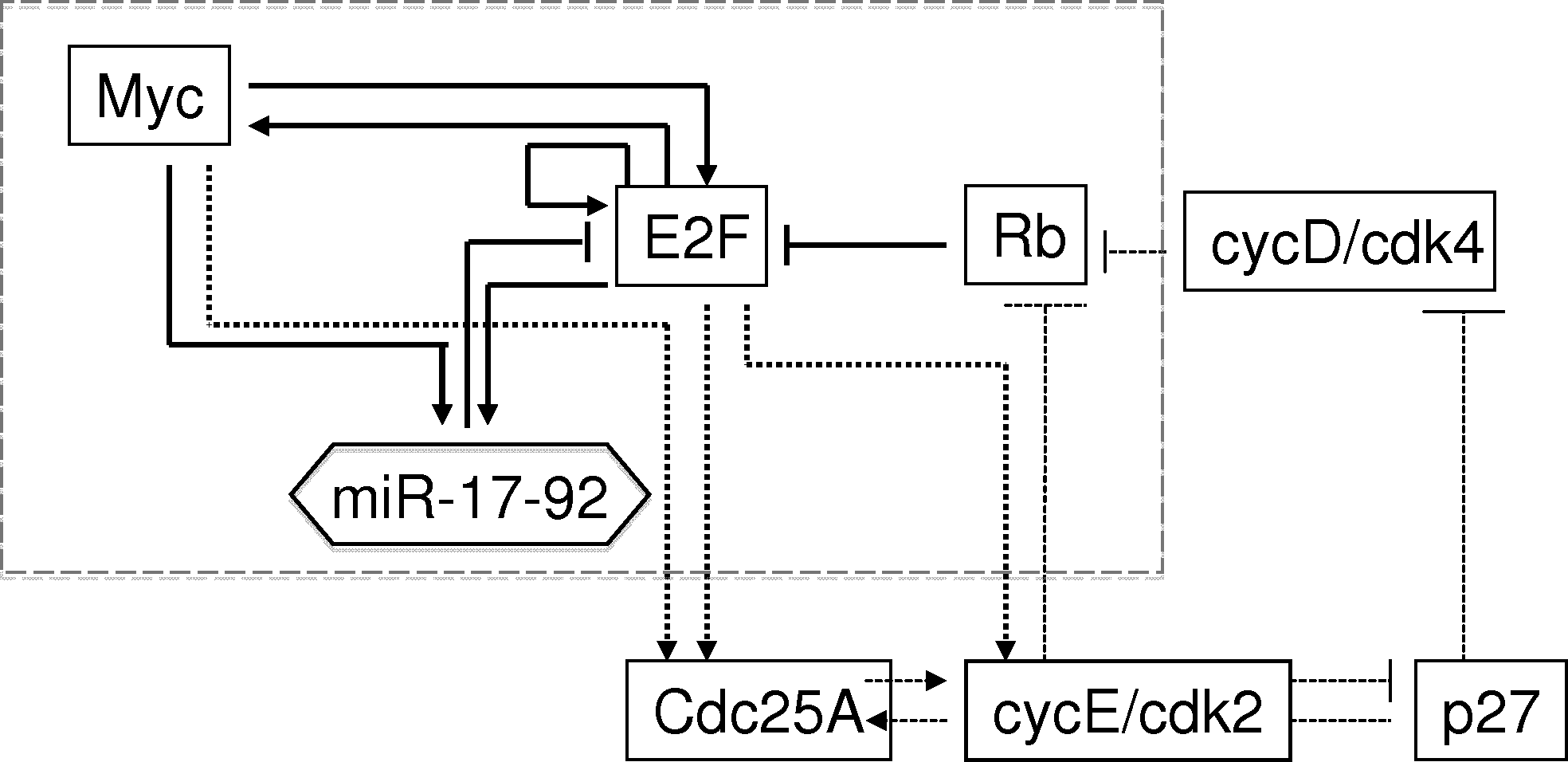}
\caption{The E2F-MYC-miR-17-92 toggle switch with its biochemical environment}
\end{figure}

Oncogenes and tumor-suppressor genes are two pivotal factors in tumorigenesis. Recent evidences indicate that MicroRNAs (miRNAs) can function as tumor suppressors and oncogenes, and these miRNAs are referred to as ‘oncomirs’. MiRNAs are small, non-coding RNAs that modulate the expression of target mRNAs.  The biogenesis pathway of miRNAs in animals was elucidated by \cite{Bartel}.  MiRNAs undergo substantial processing since the nuclear transcription where two proteins play an essential role: Drosha and Dicer. Most of miRNA are first processed into pre- miRNA by Drosha. After exportated to the cytoplasm, the pre- miRNA is processed by Dicer into a small double strand RNA (dsRNA) called the miRNA: miRNA duplex. The active strand, which is the mature miRNA is incorporated into the RISC and binds to the target mRNA, whereas the inactive strand is ejected and degraded. In normal tissue, proper regulation of miRNAs maintains a normal rate of development, cell growth, proliferation, differentiation and apoptosis. Tumorigenesis can be observed when the target gene is an oncogene, and the loss of the miRNA, which functions as a tumor suppressor, might lead to a high expression level of the oncoprotein. When a miRNA functions as an oncogene, its constitutive amplification or overexpression could cause repression of its target gene, which has a role of tumor suppressor gene, thus, in this situation, cell is likely to enter tumorigenesis. MiRNAs are often part of toggle switches, with important examples are gene pairs built with oncogenes and tumour suppressor genes \cite{Lim,Lu}. 
Here we focus on the amplification of 13q31-q32, which is the locus of the the miR-17-92. The miR-17-92 cluster forms a bistable switch with Myc and the E2F proteins\cite{ODonnell,Bueno,Aguda}.
The oncogene Myc regulates an estimated 10\% to 15\% of genes in the human genome, while the disregulated function of Myc is one of the most common abnormalities in human malignancy \cite{Coller,Remondini}. The other component of the toggle is the E2F family of transcription factors, including E2F1, E2F2 and E2F3, all driving the mammalian cell cycle progression from G1 into S phase. High levels of E2Fs, E2F1 in particular, can induce apoptosis in response to DNA damage. The toggle also interacts with dozens of genes (see figure \ref{fig:comptog} depicts a portion), particularly with the Rb and other key cell cycle players. A summary of the experiments perturbing miRNA/Myc/E2F and E2F/RB behaviours have suggested the following:
\begin{itemize}
\item The Rb/E2F toggle switch is OFF when RB inhibits E2F, i.e. stopping cell proliferation; it is ON when E2F prevails and induces proliferation. Once turned ON by sufficient stimulation, E2F can memorize and maintain this ON state independently of continuous serum stimulation.
\item
The proteins E2F and Myc facilitate the expression of each other and the E2F protein induces the expression of its own gene (positive feedback loop). They also induce the transcription of microRNA-17-92 which in turn inhibits both E2F and Myc (negative feedback loop) \cite{Sylvestre}.
\end{itemize}
Moreover, the increasing levels of E2F or Myc drive the sequence of cellular states, namely, quiescence, cell proliferation (cancer) or cell death (apoptosis).  

Although there is increasing amount of research on cell cycle regulation, the mathematical description of even a minimal portion of the E2F, Myc and miR-17-92 toggle switch is far from trivial. Aguda and collaborators \cite{Aguda} have developed a deterministic model, which reduces the full biochemical network of the toggle switch to a protein (representing the E2F-Myc compound) and the microRNA-17-92 cluster (seen as a single element).

It is a 2-dimensional open system, in which $p$ represents the E2f-myc complex and $m$ the miRNA cluster: thus no mass action law holds, i.e. $p+m\neq N$. 
The dynamics of $p$ and $m$ concentrations are described by eq. \ref{aguda1}  and \ref{aguda2}:
\begin{equation}  
\dot{p} = \alpha + \frac{k_1 \cdot p^2}{\Gamma_1 + p^2 + \Gamma_2\cdot m} -  \delta\cdot p
\label{aguda1}
\end{equation}
\begin{equation} \dot{m} = \beta + k_2\cdot p -\gamma\cdot m
\label{aguda2}
\end{equation}
The system can be rewritten in an  adimensional form as follows:
\begin{equation}  \epsilon \dot{\phi} = \alpha' + \frac{k \cdot {\phi}^2} {{\Gamma'}_1 + {\phi}^2 + {\Gamma'}_2\cdot {\mu}} - \phi
\label{aguda3}
\end{equation}
\begin{equation} \dot{\mu} = 1 + \phi - \mu
\label{aguda4}
\end{equation}
Where the parameters are:
$\alpha'=\frac{k_2}{\delta \cdot \beta} \alpha$,
$k=\frac{{k_1}{k_2}}{\delta \beta}$,
${\Gamma'_1}=\frac{k_{2}^2}{\beta^2}\Gamma_1$,
${\Gamma'_2}=\frac{k_{2}^2}{{\beta}\gamma}\Gamma_2$,
$\epsilon=\frac{\gamma}{\delta}$ and the change of variables is:
$\phi= \frac{k_2}{\beta} p$, $\mu=\frac{\gamma}{\beta}m$ and $\tau=\gamma t$.

In the original model \cite{Aguda}, the rate of protein synthesis is not a function of the instantaneous concentration (as assumed in eq.\ref{aguda3} ) but rather of its concentration at some time ${\Delta}$ in the past:
\begin{equation}  
\epsilon \dot{\phi} = \alpha' + \frac{k[{\phi}(\tau-\Delta)]^2} {{\Gamma'}_1 + [{\phi}(\tau-\Delta)]^2 + {\Gamma'}_2\cdot {\mu}(\tau-\Delta)} - \phi(\tau). 
\end{equation}
We will not consider such delay in our stochastic realization of the model, since it would increase system dimensionality and it does not seem necessary to obtain the features we want to characterize.

The steady state can be studied in the nondimensionalized system and, therefore, the conditions on the parameters for the existence of multiple steady states. In the resulting cubic equation:
\begin{equation} 
\alpha' + \frac{k{\phi}^2} {{\Gamma'}_1 + {\phi}^2 + {\Gamma'}_2\cdot (1+\phi)} - \phi = 0
\label{agudaCubic}
\end{equation}
the necessary (but not sufficient) conditions for the existence of 3 steady states (and thus a bistable system) are:
\begin{equation} 
(\Gamma'_2 - k) < \alpha' < \bigg(1+\frac{\Gamma'_1}{\Gamma'_2}\bigg) 
\end{equation}

\section{The stochastic modeling approach}

The system represented by equations \ref{aguda1} and \ref{aguda2} can studied as a stochastic system through the Chemical Master Equation (CME) approach \cite{VanKampen}.
The resulting CME has two variables, the number of $p$ and $m$ molecules, labeled as $n$ and $m$.
The temporal evolution in the probability, $p_{n,m}(t)$, to have $n$ and $m$ molecules at time $t$ is described by the following equation:
\begin{equation}
\dot{p}_{n,m} = (\mathbb E_n -1) r^n p_{nm}+ (\mathbb E^{-1}_n-1) g^n p_{nm}\\
+(\mathbb E_m -1) r^m p_{nm}+ (\mathbb E^{-1}_m-1) g^m p_{nm}
\end{equation}
This CME is derived under the condition of a one-step Poisson process, $\mathbb E $ and $\mathbb E^{-1}$
are the forward and backward step operators, $g$ and $r$ the generation and recombination terms for the $n$ and $m$ variables, as shown in superscripts.

The two generation and recombination terms associated with the $n$ and $m$ variables are respectively:
\begin{equation} 
g^n= \alpha + \frac{k_1 \cdot n^2}{\Gamma_1 + n^2 + \Gamma_2\cdot m}; \qquad r^n= \delta\cdot n \end{equation}
\begin{equation} 
g^m= \beta + k_2\cdot n; \qquad r^m= \gamma\cdot m
\end{equation}

We remark that the molecule influxes into the system (represented by the $\alpha$ and $\beta$ terms) could be included in different ways in the stochastic equations, since in the deterministic equations they represent a sort of "mean field" value. As an example, molecules could be added in bursts with specific time distributions, that do not appear in the macroscopic continuous deterministic equations. We will consider the simplest approach, but the choice of different influx patterns should deserve further investigation.

\subsection{The one-dimensional model}

We can reduce the problem from two to one dimension, by considering a different time scale for the two reactions (in particular considering $\dot{m}\gg\dot{p}$) and thus considering the steady state solution for the $m$:
\begin{equation}m=\frac{\beta+k_2\cdot p}{\gamma}=\beta'+k'\cdot p\end{equation}
As a consequence we obtain a deterministic equation for the $p$ only:
\begin{equation}\dot{p}=\alpha+\frac{k_1\cdot p^2}{\Gamma'+\Gamma''\cdot p +
p^2}-\delta\cdot p\end{equation}
Where ${\Gamma}'=\frac{{\Gamma_2} \cdot k_2}{\gamma} $ and ${\Gamma}''={\Gamma_1} + \frac{\Gamma_2 \beta}{\gamma} $.
The stochastic equation for $p_{n}$ is thus as follows:

\begin{equation}
\label{stoch1d}
\dot{p_{n}}=(\mathbb E-1) r_{n} \cdot p_{n} + (\mathbb E^{-1} -1) g_{n} \cdot p_{n} \end{equation}
\begin{equation}g_{n}= \alpha +\frac{k_1 \cdot n^2}{\Gamma'+ \Gamma''\cdot n + n^2};
\qquad r_{n}=\delta \cdot n 
\end{equation}

A general solution can be obtained
\begin{equation}
\label{GenSol}
p_{n}^s=\prod_{i=1}^{N} \frac{g(i-1)}{r(i)} \cdot p_0=\prod_{i=1}^{N} \frac{\alpha + \frac{k_1 \cdot
(i-1)^2}{\Gamma'+\Gamma'' \cdot (i-1)+(i-1)^2}} {\delta \cdot i} \cdot p_0
\end{equation}

with an adequate normalization factor imposed on $p_0$:
\begin{equation} 
p_0=\frac{1}{1+\sum_{i=1}^{N}\prod_{i=1}^{N}{p_{n_1}^s}}
\end{equation}
We remark that the system is open, thus in theory $N$ is not fixed, but we can truncate the product to a sufficiently high value of $N$ obtaining a good approximation of the whole distribution. 
This one-dimensional system (for which an analytical solution can be obtained) will be compared to numerical simulations of the exact one-dimensional and two-dimensional systems.

\section{Model Analysis}

\subsection{The stationary distribution}

\begin{figure}[th]
\centerline{\includegraphics[height=50 mm]{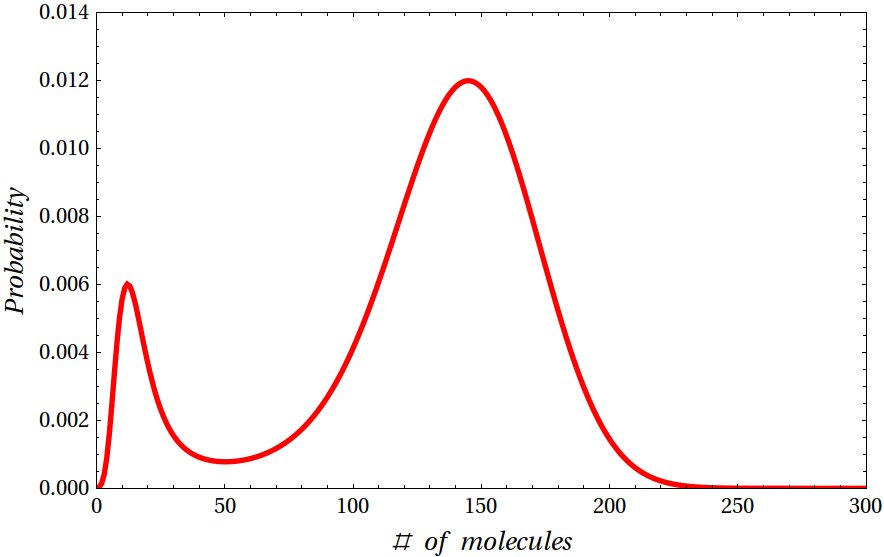}}
\caption[]{The stationary distribution for the one-dimensional space, obtained using the following parameters: $\alpha=1.68(molecule/h)$, $\beta=0.202(molecule/h)$, $\delta=0.2 (h^{-1})$, $\gamma=0.2(h^{-1}) $, $\Gamma_1=10300 (molecule^2)$, $ {\Gamma_2}=1006(molecule)$, $k_1=90(molecule/h)$  and $k_2=0.05(h^-1)$.}
\label{Staz_Distr_fig}
\end{figure}

The one-dimensional model can show monomodal as well as bimodal stationary distributions, depending on the parameters considered. As an example, we obtain bistability with a set of parameters as shown in Fig. \ref{Staz_Distr_fig}. 

Thus the qualitative features of the two-dimensional deterministic model (i.e. the possibility of being bistable depending on the parameter range) are recovered  for the one-dimensional approximation of the stochastic system. Also the two-dimensional stochastic system shows bistability for the same parameters, and they are in optimal agreement for a range of parameters in which the $\dot{m}\gg\dot{p}$ condition holds \\

\begin{figure}\centering
\includegraphics[width=0.5\textwidth]{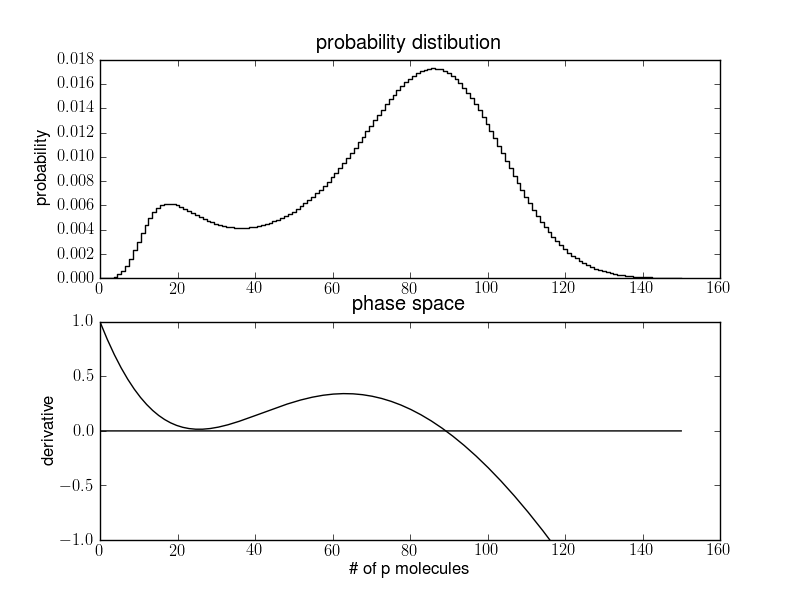}
\caption{Comparison between the deterministic solution (bottom) and the stationary distribution (top) for the parameter set as in Table \ref{Table}, case 3.}
\label{ghostFig1} 
\end{figure}

We also observe some remarkable differences between the deterministic and the stochastic models: there are regions in parameter space in which the deterministic approach shows only one stable state, but in the stochastic system two maxima in the stationary distribution are observed (see Fig. \ref{ghostFig1}).
This difference can be explained qualitatively as follows: for the deterministic system, there are parameter values for which the system is monostable but very close to the "transition point" in which the system becomes bistable. It is known that in these situations a "ghost" remains in the region where the stable point has disappeared \cite{Strogatz}, for which the systems dynamics has a sensible slowing down (i.e. when the system is close to the disappeared fixed point, it remains "trapped" for a longer time close to it, in comparison with other regions). This behaviour results in the presence of a peak in the stationary distribution of the corresponding stochastic systems, that thus remains bistable also when the deterministic system is not anymore. 

\begin{figure}\centering
\includegraphics[width=0.5\textwidth]{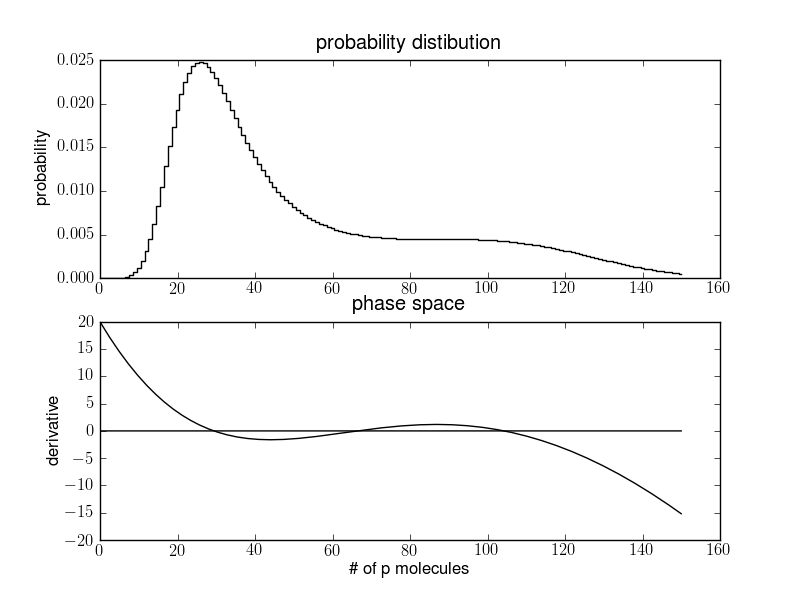}
\caption{Comparison between the deterministic solution (bottom) and the stationary distribution (top) for the parameter set as in Table \ref{Table}, case 4.}
\label{ghostFig2} 
\end{figure}

Another difference is observed: for some parameter values the deterministic system is bistable, but the stochastic distribution shows a clear peak for the maximum with the largest basin of attraction and the smaller peak results ''masked'' by the tail of the distribution around the first peak (see Fig. \ref{ghostFig2}), thus resulting in a monomodal distribution with a long tail.
In practice, the highest state behaves like a sort of metastable state, since the states of the system with a high protein level are visited only occasionally. 

\subsection{Numerical analysis}

Here we implemented numerical methods to find the stationary distribution of a CME. The most accurate is the Kernel resolution method: given the complete transition matrix of the system, it  is possible to solve numerically the eigenvalue problem, obtaining the correct stationary distribution. 
This method, in this case, has a serious drawback: the system is of non-finite size, preventing a complete enumeration of the possible states. Even with a truncation, the system size rises in a dramatic way: the state space for a bidimensional system is of order $N^2$ if $N$ is the truncation limit, and thus the respective transition matrix is of order $N^4$. 
This means that even for a relatively small system (with a few hundred of molecules) the matrix size explodes well beyond the computational limits. 
The only feasible resolution strategy is a massive exploration of state space by Montecarlo methods, in which single trajectories of the system are simulated: performing this simulations long enough for several times allows to estimate the stationary distribution.

The Montecarlo method we chose is a modified version of the SSA algorithm (also known as the Gillespie algorithm) named logarithmic direct method \cite{Gillespie,Li}, which is a statistically correct simulation of an ergodic Markov system. It is not the fastest algorithm available, as compared to other methods like the next-reaction or the $\tau$-leap method, but it produces a correct estimation of the statistical dispersion of the final state.

For each parameter set  we performed 10 simulations for about $10^6 - 10^7$ iteration steps each.
The multiple simulations were averaged together for a better estimation of the stationary distribution, and they allowed also an estimation of the variance over this average distribution.



In the following we discuss four cases that describe the system behaviour for different parameter settings, shown in Table \ref{Table}.

\begin{table}
\caption{Table of the parameter sets for the cases considered.}
\begin{tabular}{|l||c|c|c|c|}
\hline
Par&Case 1&Case 2&Case 3&Case 4\\
\hline
$\alpha$ (molecule/h) &1.0&1.68&1.0&20.0\\
$\delta$ $(h^{-1})$ &1.0&0.20&0.09&1.19\\
$\beta$	(molecule/h) &1.0&0.202&0.0&1.0\\
$\gamma$ $(h^{-1})$ &100.0&0.20&10.0&1.0\\
$k_1$ (molecule/h) &30.0&90.0&12.5&230.0\\
$k_2$ $(h^-1)$ &100.0&0.05&10.0&1.0\\
$\Gamma_1$ (molecule$^2$)	&60.0&10300.0&$(72.8)^2$&$(110.0)^2$\\
$\Gamma_2$ (molecule) &10.0&1006.0&10.0&10.0\\
\hline
\end{tabular}
\label{Table}
\end{table}

\begin{figure}
\includegraphics[width=0.5\textwidth]{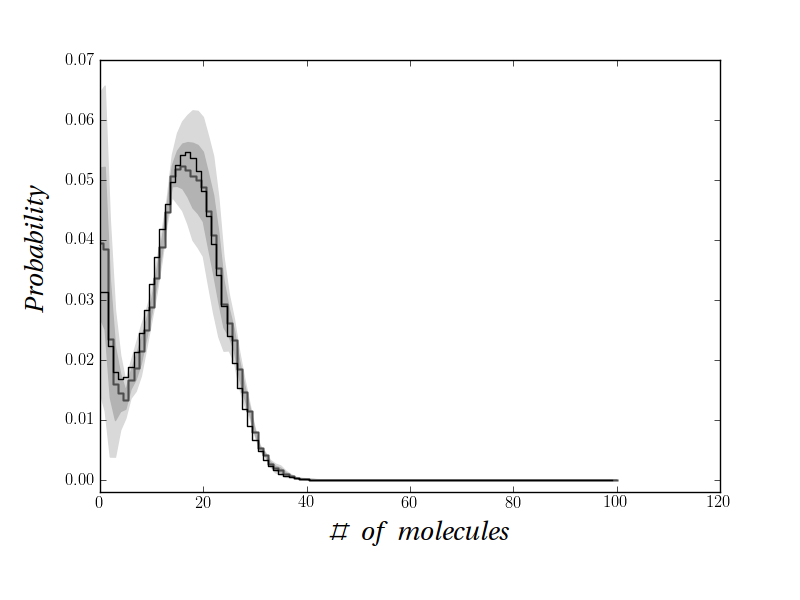}\includegraphics[width=0.5\textwidth]{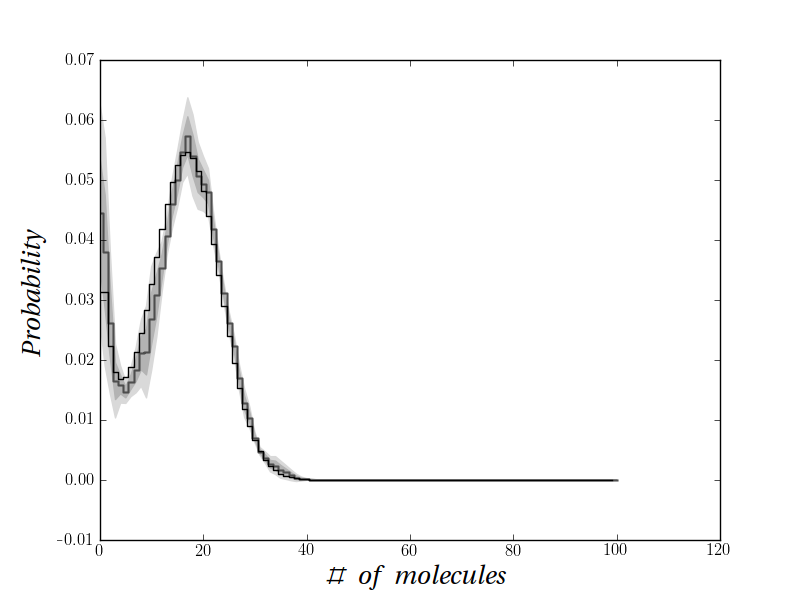}
\caption{Case of good agreement between the theoretical and obtained distribution (see Tab. \ref{Table}, case 1). Left: one-dimensional system, right: two-dimensional system. The thin black line is the theoretical distribution obtained from Eq. \ref{GenSol}. The thick dark grey line is the average of the various simulations, while the grey and light grey areas represent the range of one and two standard deviations from the average distribution.}
\label{Agreement}
\end{figure}

In case 1, we have a system in which the hypothesis of a time-scale separation between $m$ and $p$ is strongly satisfied.
The simulation was performed up to a time limit of $10^3$: we can see how the two resulting distributions are in good agreement with the theoretical one (see Fig. \ref{Agreement}), with the regions of higher variance of the histogram around the maxima and minima of the distribution.

\begin{figure}
\includegraphics[width=0.5\textwidth]{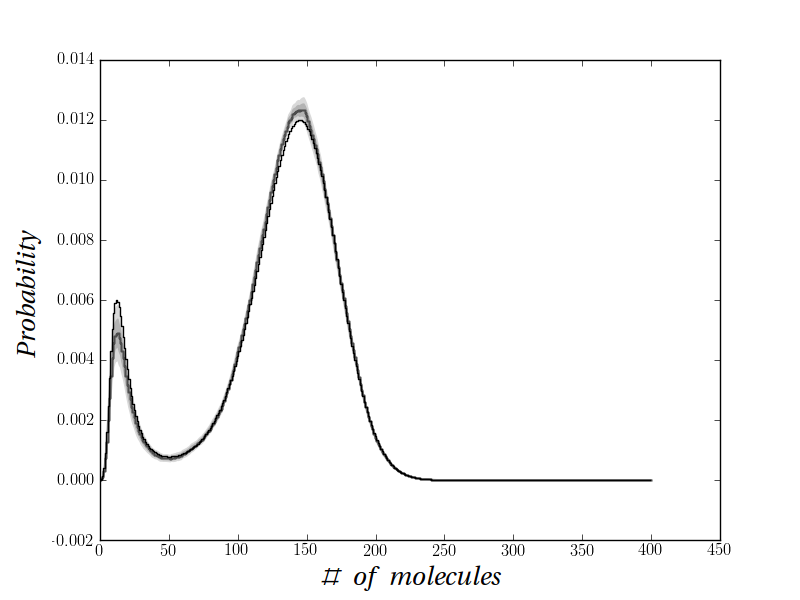}\includegraphics[width=0.5\textwidth]{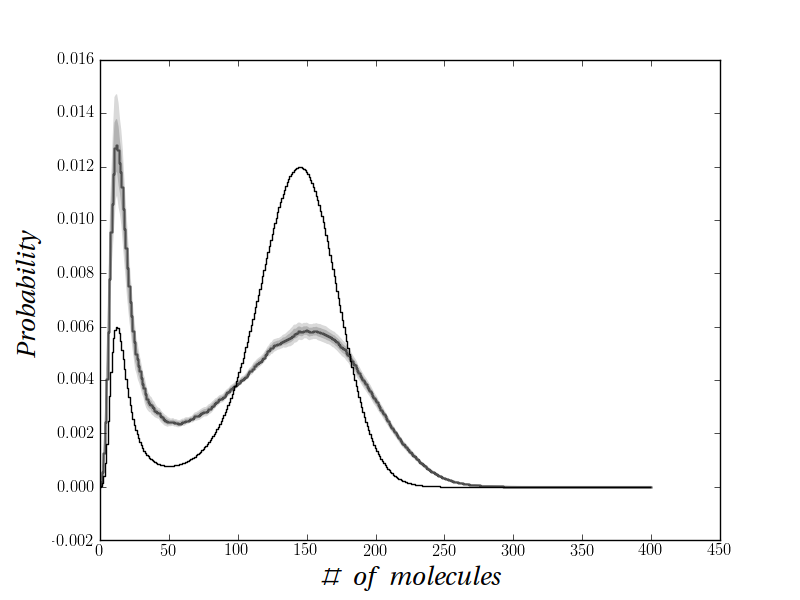}
\caption{Case of poor agreement between the theoretical and obtained distribution (see Tab. \ref{Table}, case 2). Left: one-dimensional system, right: two-dimensional system. The thin black line is the theoretical distribution obtained from Eq. \ref{GenSol}. The thick dark grey line is the average of the various simulation, while the grey and light grey areas represent the range of one and two standard deviations from the average distribution.}
\label{Disagreement}
\end{figure}

In case 2, the time-scale separation assumption does not hold, due to the very low value of $\gamma$ and $k_2$: even if this condition doesn't guarantee that the stationary state will be different from the approximate one-dimensional solution, with this set of parameters we can see a huge difference between the two distributions (Fig. \ref{Disagreement}).

\begin{figure}
\includegraphics[width=0.5\textwidth]{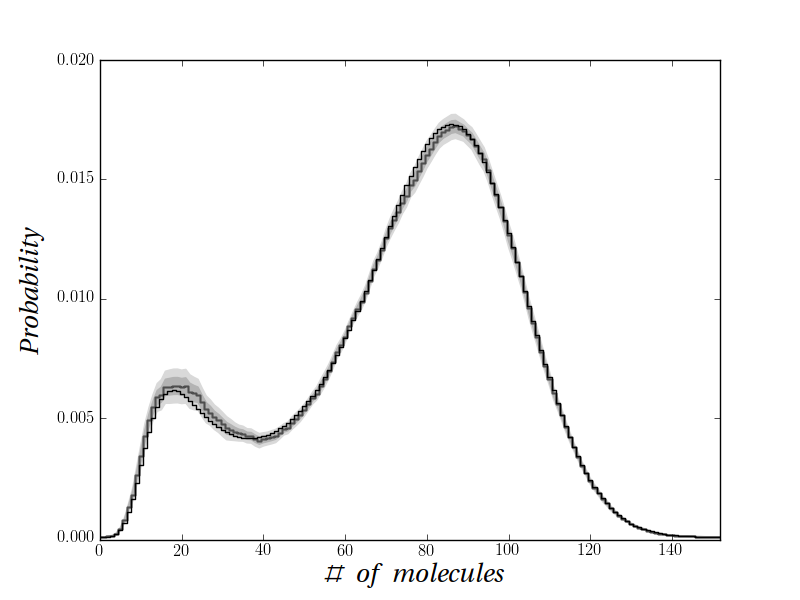}\includegraphics[width=0.5\textwidth]{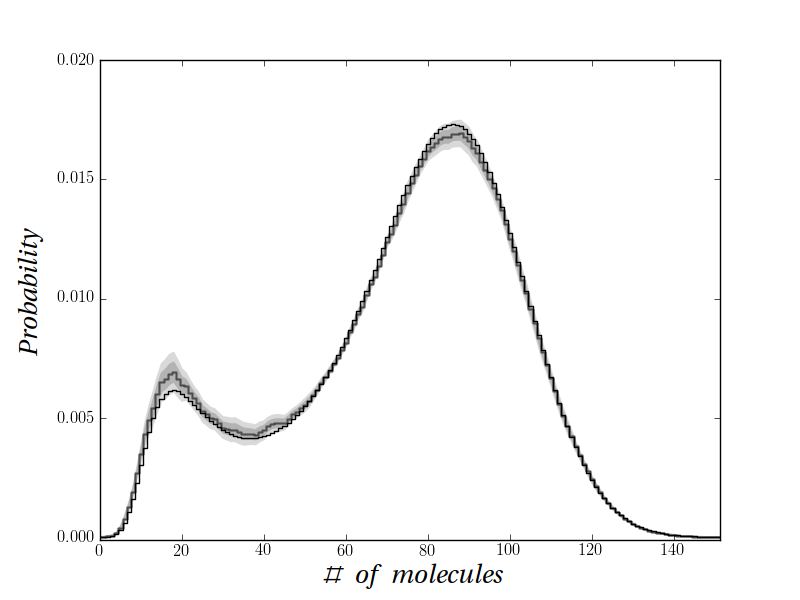}
\caption{Case 3, ''ghost effect'': only the biggest peak comes from a deterministic stable point. Left: one-dimensional system, right: two-dimensional system. The thick dark gray line is the average of the various simulation, while the gray and light gray areas represent the range of one and two standard deviations from the average distribution.}
\label{Ghost}
\end{figure}

In case 3, as defined before, we observe a ''ghost'' in which, even if a deterministic stable state does not exist, we can clearly see a second peak in the distribution (Fig. \ref{Ghost}).
In this system the time-scale separation assumption holds, and we can see how both distributions show similar features. 

\begin{figure}
\includegraphics[width=0.5\textwidth]{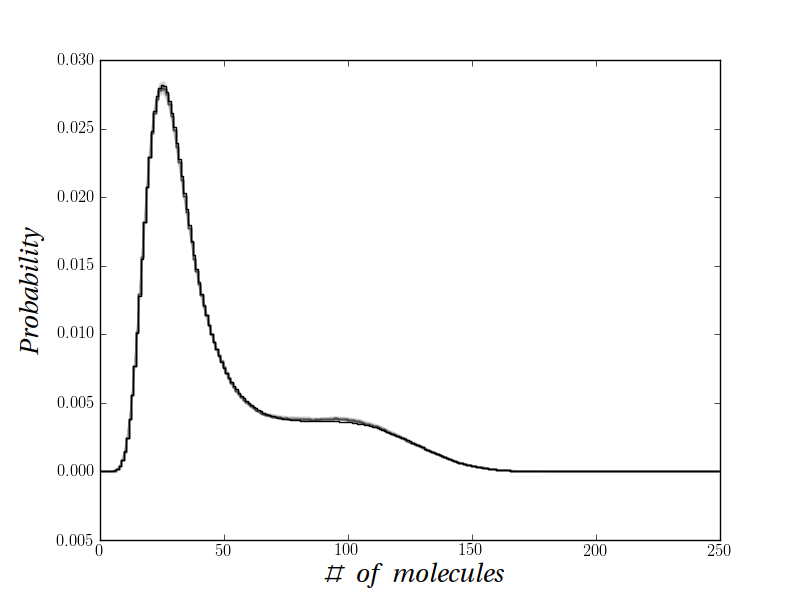}\includegraphics[width=0.5\textwidth]{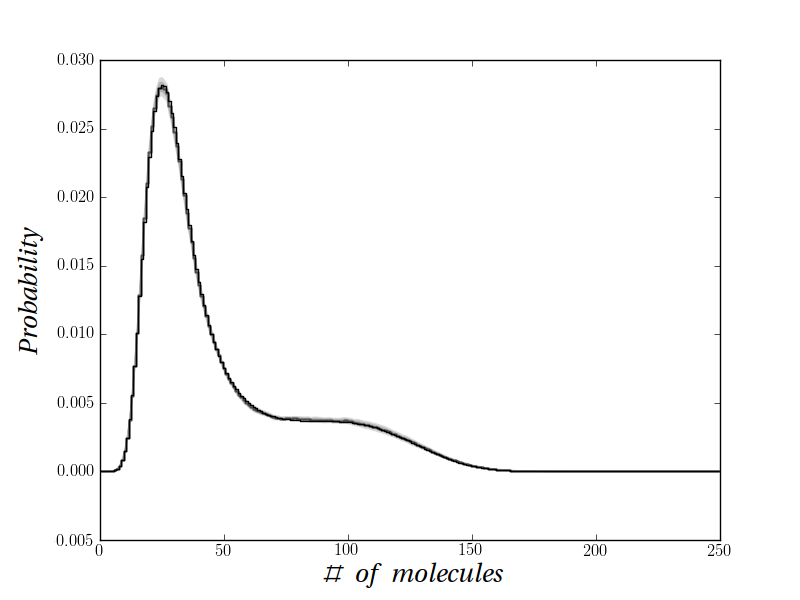}
\caption{Case 4, peak masking effect (parameters as in Tab. \ref{Table}, case 4). The deterministic system has two stable points, but only the peak related to the smallest stable point (with the largest basin of attraction) is visible. Left: one-dimensional system, right: two-dimensional system.}
\label{MaskedPeak}
\end{figure}

In this final case (Tab. \ref{Table}, case 4, Fig. \ref{MaskedPeak}) we can see another effect, in which the peak related to a deterministic stable state is masked by the tail of the stronger peak, becoming just a fat tail.
Even without a strong time-scale separation for the $m$ and $p$ variables, we can see how both systems give a very similar response, evidencing that this effect is very robust.
Increasing the $\gamma$ and $k_2$ values does not affect the distribution as long as their ratio is kept constant.
Note that while there are several computational tools for discrete-state Markov processes such as PRISM \cite{Kwiatkowska}, APNNtoolbox \cite{Buchholz}, SHARPE \cite{Hirel},  or Mobius \cite{Daly}, there is very little for CMTC (see for instance \cite{Didier}).  Different modeling approaches for toggle switches do exists in the area of formal methods (see for example \cite{Bella1,Bella2}).

\section{Discussion and Conclusion}

We have studied a stochastic version of a biochemical circuit that is supposed to be involved in cell cycle control, with implications for the onset of severe diseases such as cancer, consisting of a gene cluster (Myc-E2F) and a miRNA cluster (mir-17-92). This cluster  has been reported in very large number of cancer types: particularly in different types of lymphomas, glioma, non-small cell lung cancer, bladder cancer, squamous-cell carcinoma of the head and neck, peripheral nerve sheath tumor, malignant fibrous histiocytoma, alveolar rhabdomyosarcoma, liposarcoma and colon carcinomas. This huge variety of cancer stresses the centrality of this toggle switch and suggests that advancement in modeling this toggle could lead to insights into differences between these cancers. This aim is still far but we are delighted to report that our modeling approach shows important results inching to that direction. 
First of all, many features are recovered as observed for the deterministic version of the same system, also by means of a further approximation that reduces the system to an unique variable: in this case the system can be treated analytically, and compared to the one- and two-dimensional numerical simulations.

The stochastic approach, that is the exact approach when the number of molecules involved is low, shows a different behaviour than the deterministic one in two situations we have observed. It is noteworthy that the number of molecules involved shows some agreement with the estimates by \cite{Chan} and by \cite{Lim2} for other miRNA-systems (see also \cite{Arvey}). The cell volume is assumed  $10^{-13}$ liters, then 1 nM =100 molecules.

First, bistability in the stochastic system (namely, the possibility of having two stable states, one associated to a resting and the other to a proliferative cell state) is observed also in situations in which the corresponding deterministic system is monostable, and this can be explained by the presence of a ''ghost'' state in the deterministic system that is strong enough to produce a second peak in the stationary distribution of the stochastic model. 

Secondly, there are situations in which the peak for the stochastic distribution related to the highest level of expression (with parameter values for which the deterministic system is bistable) is masked by the tail of the distribution of the lowest-expression maximum (that is related to the largest basin of attraction in the deterministic model), making  the ''proliferative state'' appear almost as a scarcely visited metastable state. This is an interesting behaviour, that should be further investigated in real experimental data of protein concentration and gene expression related to the biochemical circuit considered. The ''metastable'' and the ''fully'' bimodal distributions could be associated to healthy and tumoral cell states respectively, because the highest ''proliferative'' state has different properties in the two cases. From a biological point of view such state, being associated to a dysregulated, disease-related conditions, could actually represent a compendium of several dysregulated states. 

We argue that the deterministic approach to this biochemical circuit is not capable to characterize it completely, and the stochastic approach appears more informative: further features unique to the stochastic model could be obtained by considering different time patterns for the molecular influxes to the system, and this point in our opinion should deserve more investigation in a future work. MicroRNAs (miRNAs) express differently in normal and cancerous tissues and thus are regarded as potent cancer biomarkers for early diagnosis. We believe that the potential use of oncomirs in cancer diagnosis, therapies and prognosis will benefit accurate cancer mathematical models. 

Given that MiR-17-5p seems to act as both oncogene and tumor suppressor through decreasing the expression levels of anti-proliferative genes and proliferative genes, this behavior is suggestive of a cell type dependent toggle switch. Therefore fitting of experimental data could provide insights into differences among cancer types and on which cell type is behaving differently.


\section*{Acknowledgments} D.R. acknowledges Bologna University 'Progetto Strategico d'Ateneo 2006' funding.

\end{document}